\documentclass[12pt]{iopart}

\usepackage{iopams} 
\usepackage{verbatim}
\usepackage{graphicx}
\usepackage{color}
\begin{document}

\title{Reconfigurable Stochastic Neurons Based on Strain Engineered Low Barrier Nanomagnets}

\author{Rahnuma Rahman, Samiran Ganguly and Supriyo Bandyopadhyay}

\address{Department of Electrical and Computer Engineering, Virginia Commonwealth University, Richmond, VA 23284, USA}
\ead{sbandy@vcu.edu}
\vspace{10pt}

\begin{abstract}
Stochastic neurons are efficient hardware accelerators for solving a large variety of combinatorial optimization problems. ``Binary'' stochastic neurons (BSN) are those whose states fluctuate randomly between two levels +1 and -1, with the probability of being in either level  determined by an external bias. ``Analog'' stochastic neurons (ASNs), in contrast, can assume any state between the two levels randomly (hence ``analog'') and can perform analog signal processing. They may be leveraged for such tasks as temporal sequence learning, processing and prediction. Both BSNs and ASNs can be used to build efficient and scalable neural networks. Both can be implemented with low (potential energy) barrier nanomagnets (LBMs) whose random magnetization orientations encode the binary or analog state variables. The difference between them is that the potential energy barrier in a BSN LBM, albeit low, is much higher than that in an ASN LBM. As a result, a BSN LBM has a clear {\it double well potential profile}, which makes its magnetization orientation assume one of two orientations at any time, resulting in the binary behavior. ASN nanomagnets, on the other hand,  hardly have any energy barrier at all and hence lack the double well feature. That makes their magnetizations fluctuate in an analog fashion. Hence, one can reconfigure an ASN to a BSN, and vice-versa, by simply raising and lowering the energy barrier.  If the LBM is {\it magnetostrictive}, then this can be done with local (electrically generated) strain. Such a reconfiguration capability heralds a powerful field programmable architecture for a p-computer, and the energy cost for this type of reconfiguration is miniscule. There are also other applications of strain engineered barrier control, e.g.,  adaptive annealing in energy minimization computing (Boltzmann or Ising machines),  emulating memory hierarchy  in a dynamically reconfigurable fashion, and control over belief uncertainty in analog stochastic neurons. Here, we present a study of this modality.

\end{abstract}

\medskip


%
%
%
\maketitle
%
%

\section{Introduction: Binary and analog stochastic neurons implemented with LBMs }

Binary stochastic neurons (BSNs) are a well known route to implementing ``spins'' in Ising machines \cite{orchi1, orchi2} and have been used to solve computationally hard problems such as graph theoretic problems like Max Cut \cite{khilwani}, factorization \cite{borders}, etc. very efficiently. A popular approach to realizing them is with a low (energy) barrier nanomagnet (LBM), whose magnetization fluctuates randomly between two preferred orientations representing the binary bits +1 and -1. The probability of being in either bit can be altered by a ``bias'', such as a spin-polarized current injected into the LBM \cite{orchi1}. 

\begin{figure}[h!]
\centering
\includegraphics[width=6in]{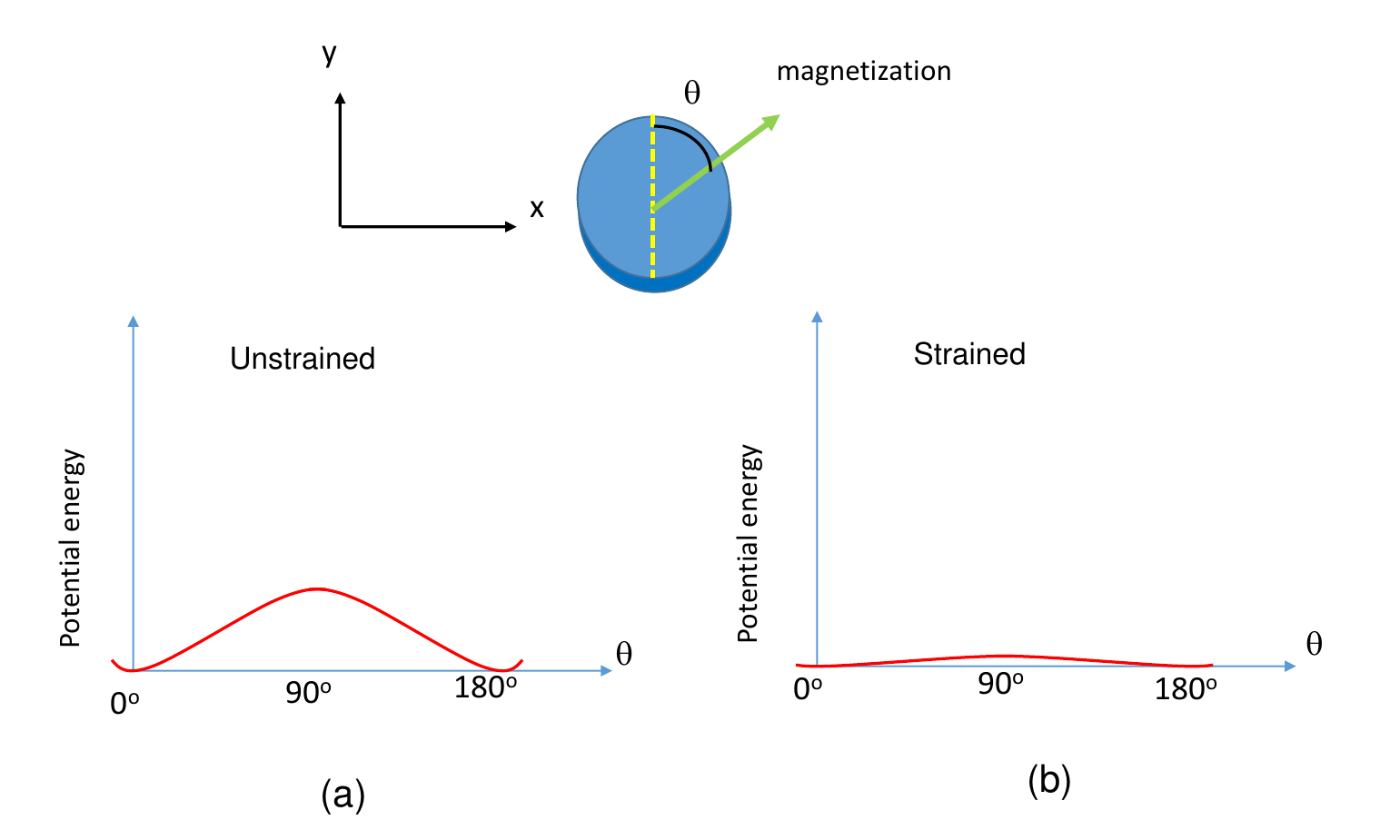}
    \caption{\small (a) Potential energy as a function of the in-plane magnetization orientation in a magnetostrictive nanomagnet shaped like an elliptical disk with small eccentricity. (b) Potential energy as a function of the in-plane magnetization orientation when the nanomagnet is subjected to uniaxial stress along the major axis such that the sign of the product of the stress and the magnetostriction is negative. The inset shows the nanomagnet and the magnetization orientation, with $\theta$ being the angle subtended by the magnetization with the nanomagnet's major axis.}
\label{fig:potential}
\end{figure}

The LBM is usually a nanomagnet with in-plane anisotropy that is shaped like an elliptical disk with small (but non-zero) eccentricity. The in-plane potential energy profile (energy versus magnetization orientation) of such a LBM is shown schematically in Fig. \ref{fig:potential}(a). Normally, there is a clear double-well feature which can be discerned despite the low potential barrier. The two ground states (or wells) correspond to the magnetization pointing along either direction along the major axis (or easy axis) of the elliptical nanomagnet. At room temperature, thermal energy can allow the magnetization to transcend the energy barrier separating the wells, which will allow the magnetization to fluctuate randomly between the two potential wells. If we take a snapshot in time, we will usually find the magnetization in one of the two wells, i.e., it will tend to point along one of the two directions along the major axis, which encode the bits +1 and -1. This leads to the digital or ``binary'' behavior.

One can depress the energy barrier separating the two wells with (electrically generated) mechanical strain if the nanomagnet is made of a magnetostrictive material like Co or FeGa or Terfenol-D. For this to happen, the sign of the product of the magnetostriction and the strain has to be negative. When the energy barrier is depressed sufficiently, it begins to lose the double-well feature, as shown in Fig. \ref{fig:potential}(b). At that point, the magnetization has very little tendency to settle into either of the two degenerate ground states preferentially, and all orientations are almost equally likely. Consequently, the magnetization will fluctuate among all orientations with the same likelihood, leading to the ``analog'' behavior. We  can, therefore, {\it reconfigure} a stochastic neuron from binary to analog, and vice-versa, with strain.

\section{Strain effects on energy barrier}

The steady-state in-plane potential energy in a strained elliptical nanomagnet with in-plane anisotropy depends on the magnetization orientation as \cite{fashami}
\begin{equation}
  E  = \left ( \mu_0/2 \right ) M_s^2 \Omega \left [ N_1 cos^2 \theta + N_2 sin^2 \theta \right ] - (3/2) \lambda_s Y \epsilon \Omega cos^2 \theta,
  \label{energy}
\end{equation}
where $\mu_0$ is the permeability of vacuum, $M_s$ is the saturation magnetization of the nanomagnet's material, $\Omega$ is the nanomagnet volume, $\lambda_s$ is the saturation magnetostriction of the nanomagnet material, $Y$ is the Young's modulus of the nanomagnet, $\epsilon$ is the strain, $\theta$ is the angle shown in the inset of Fig. \ref{fig:potential} to denote the magnetization orientation (it is the angle subtended by the in-plane component of the magnetization with the major axis of the elliptical nanomagnet), and 
\begin{eqnarray*}
N_1 & = & {{\pi\over{4}}} \left ( {{t}\over{a}} \right )\left [1 - \frac{1}{4} \left ( \frac{a-b}{a} \right ) - \frac{3}{16} \left ( \frac{a-b}{a} \right )^2 \right ] \\
N_2 & = & {{\pi\over{4}}} \left ( {{t}\over{a}} \right ) \left [1 + \frac{5}{4} \left ( \frac{a-b}{a} \right ) + \frac{21}{16} \left ( \frac{a-b}{a} \right )^2 \right ],
    \end{eqnarray*}
where $a$ is the major axis, $b$ is the minor axis and $t$ is the thickness of the nanomagnet \cite{chikazumi}.

\begin{figure}[h!]
\centering
\vspace{-40pt}
\includegraphics[width=6in,angle=-90]{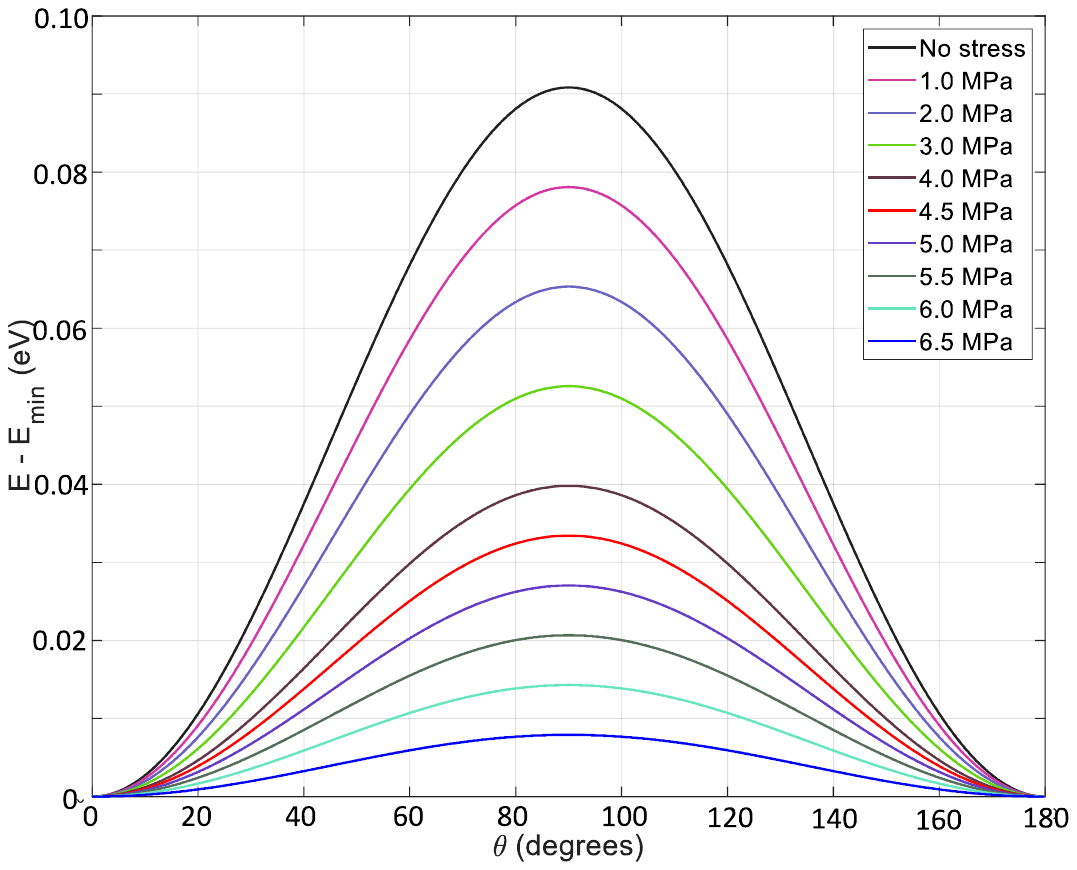}
\vspace{-40pt}
    \caption{\small The potential energy profile in a Co nanomagnet shaped like an elliptical disk as function of the angle $\theta$ subtended by the magnetization with the major axis. The results are shown for different stress values. The quantity $E$ is calculated from Equation (1) and $E_{min}$ is the minimum value of $E$. The nanomagnet has major axis = 100 nm, minor axis = 99 nm and thickness = 5 nm. }
\label{fig:energy-barrier}
\end{figure}

Equation (\ref{energy}) clearly shows that if the product $\lambda_s \epsilon$ has a negative sign, then application of strain will depress the energy barrier, which is what we depicted schematically in Fig. \ref{fig:potential}. A material like Co has negative magnetostriction and hence a {\it tensile} uniaxial strain along the major axis of the nanomagnet will depress the energy barrier. A material like FeGa or Terfenol-D has positive magnetostriction and hence a {\it compressive} uniaxial strain along the major axis will depress the energy barrier. Fig. \ref{fig:energy-barrier} shows the potential energy $E- E_{min}$ ($E_{min}$ is the minimum value of $E$) as a function of the magnetization orientation in a Co nanomagnet of major axis 100 nm, minor axis 99 nm and thickness 5 nm. The energy barrier is the maximum of this quantity, and can be seen to decrease with increasing stress.

\begin{figure}[!ht]
\centering
\includegraphics[width=6in]{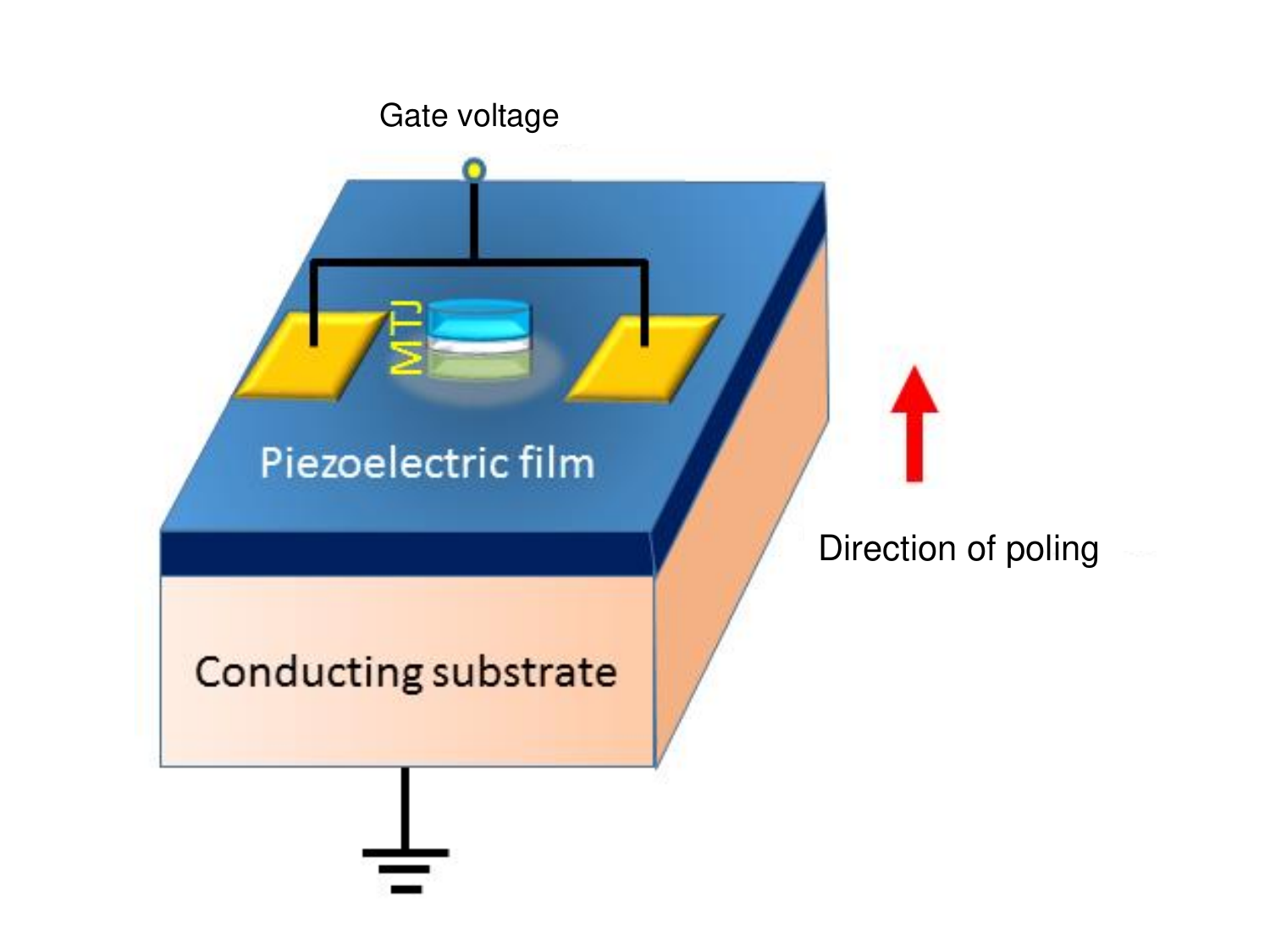}
    \caption{\small Methodology to reconfigure a BSN into an ASN and  vice versa. Applying a gate voltage of the right polarity will generate the right type of biaxial strain in the poled piezoelectric region underneath the nanomagnet. This strain will be transferred to the nanomagnet and it will lower the energy barrier in the latter, making its magnetization fluctuate in an analog manner rather than a binary manner. We can build a magnetic tunnel junction (MTJ) on top of the nanomagnet which will act as its soft layer. This MTJ will be a fluctuating resistor that can be transformed from a BSN to an ASN by turning on the gate voltage to generate strain.   }
\label{fig:strain}
\end{figure}

One way to strain a nanomagnet electrically is to use the configuration shown is Fig. \ref{fig:strain}. The nanomagnet is deposited on a poled piezoelectric film and gate pads are delineated around it such that the line joining the pads passes through the major axis. The two pads are shorted together and a voltage is applied between the shorted pads and the grounded conducting substrate. The substrate is made `conducting' so that the applied gate voltage drops mostly across the piezoelectric layer and not the substrate. If the resulting electric field is parallel to the direction of the poling, then tensile strain will appear along the major axis and compressive along the minor axis of the elliptical nanomagnet. Reversing the polarity of the gate voltage will reverse the signs of the strains. The dimensions of the nanomagnet and the electrodes, the separation between the nanomagnet edge and the nearest electrode, and the piezoelectric film thickness have to satisfy certain conditions for the biaxial strain generation as described \cite{cui}, but these conditions are relatively easy to fulfill. 

We can build a magnetic tunnel junction (MTJ) on top of the nanomagnet, with the latter acting as the soft layer whose magnetization fluctuates. This will transform the MTJ into a fluctuating resistance that acts as either a BSN (no gate voltage applied to cause strain and lower the energy barrier) or an ASN (gate voltage of the right polarity applied to cause strain that lowers the energy barrier). This is the basis of a {\it reconfigurable stochastic neuron} (RSN).

An important question now is whether the strain generated by the gate voltage can be {\it non-volatile}. This will allow the reconfiguration to be non-volatile as well. There are many reports of non-volatile remanent strain in piezoelectrics at room temperature \cite{yang, yu, lupascu, wu1, wu2, liu, bandy} although the strain's longevity has not been studied. If the strain remains non-volatile, we can reconfigure a BSN to an ASN and the reconfiguration will survive subsequent removal of the gate voltage. To revert the ASN back to a BSN, we can simply apply strain of the opposite sign, which will raise the energy barrier back in the nanomagnet and convert the ASN to a BSN.

\section{Landau-Lifshitz-Gilbert simulations to study random magnetization dynamics in an LBM under different strains}

We carried out Landau-Lifshitz-Gilbert (LLG) simulations of the magnetization dynamics in an LBM at room temperature under different strains to see how the magnetization fluctuation behaves. The LBM we studied is an elliptical Co nanomagnet of major axis 100 nm, minor axis 99 nm and thickness 5 nm. A nanomagnet of these dimensions are likely to be monodomain and hence the macrospin approximation holds. The saturation magnetization $M_s$ = 10$^6$ A/m, the magnetostriction coefficient $\lambda_s$ = -35 ppm and the Gilbert damping coefficient $\alpha$ = 0.01 correspond to a Co nanomagnet. The coupled LLG equations governing the temporal evolutions of the scalar components of the magnetization were solved with finite difference method \cite{rahnuma1, rahnuma2} with a time step of 0.1 ps. We assumed positive (tensile) uniaxial strain applied along the major axis of the nanomagnet since Co has negative magnetostriction. This will depress the energy barrier within the nanomagnet. The initial condition was that the magnetization was aligned close to the major axis of the nanomagnet.   

The coupled LLG equations describing the temporal evolution of  the three components of the magnetization are:
\setlength{\mathindent}{0.3cm}
\begin{eqnarray}
    {{d m_x(t)}\over{dt}} & = & - \gamma \left [H_z(t)m_y(t) - H_y(t)m_z(t)  \right ] \nonumber \\
    && -\alpha \gamma \left [H_y(t)m_x(t)m_y(t) - H_x(t)m_y^2(t) - H_x(t)m_z^2(t) + H_z(t)m_x(t)m_z(t) \right ] \nonumber \\
    {{d m_y(t)}\over{dt}} & = & - \gamma \left [H_x(t)m_z(t) - H_z(t)m_x(t)  \right ]  \nonumber \\
    && -\alpha \gamma \left [H_z(t)m_y(t)m_z(t) - H_y(t)m_z^2(t) - H_y(t)m_x^2(t) + H_x(t)m_x(t)m_y(t) \right ] \nonumber \\
    {{d m_z(t)}\over{dt}} & = & - \gamma \left [H_y(t)m_x(t) - H_x(t)m_y(t)  \right ] \nonumber \\
    && -\alpha \gamma \left [H_x(t)m_z(t)m_x(t) - H_z(t)m_x^2(t) - H_z(t)m_y^2(t) + H_y(t)m_y(t)m_z(t) \right ] \nonumber \\
\end{eqnarray}
where $\alpha$ is the Gilbert damping factor of the nanomagnet material, $\gamma$ is the gyromagnetic factor (a constant), $m_i(t)$ is the $i$-th component of the magnetization at time $t$, and $H_i(t)$ is the $i$-th component of the effective magnetic field experienced by the nanomagnet at time $t$. The major axis of the nanomagnet is along the y-direction and the minor axis is along the x-direction.

The effective magnetic field components are given by 
\setlength{\mathindent}{2.5cm}
\begin{eqnarray}
    H_x(t) & = & -M_s N_2m_x(t) + h_x^{noise}(t) \nonumber \\
    H_y(t) & = & -M_s N_1m_y(t) + h_y^{noise}(t) + {{3}\over{\mu_0 M_s}} \lambda_s \epsilon Y m_y (t) \nonumber \\
    H_z(t) & = & -M_s N_3m_z(t) + h_z^{noise}(t) 
\end{eqnarray}
where $N_3$ = 1 - $N_1$ - $N_2$ and $h_i^{noise}(t) = \sqrt{{{2 \alpha kT}\over{\gamma \left (1 + \alpha^2 \right ) \mu_0 M_s \Omega \Delta t}}}G_{(0,1)}^i (t)$ with $G_{(0,1)}^i (t)$ ($i = x, y, z$) being three uncorrelated  Gaussians of zero mean and unit standard deviation, $\Omega$ is the nanomagnet volume, $Y$ is the Young's modulus of the nanomagnet, $\epsilon$ is the magnitude of the strain, and $\Delta t$ is the attempt period which is the time step of the simulation.

\begin{figure*}[!ht]
    \includegraphics[width=0.46\textwidth]{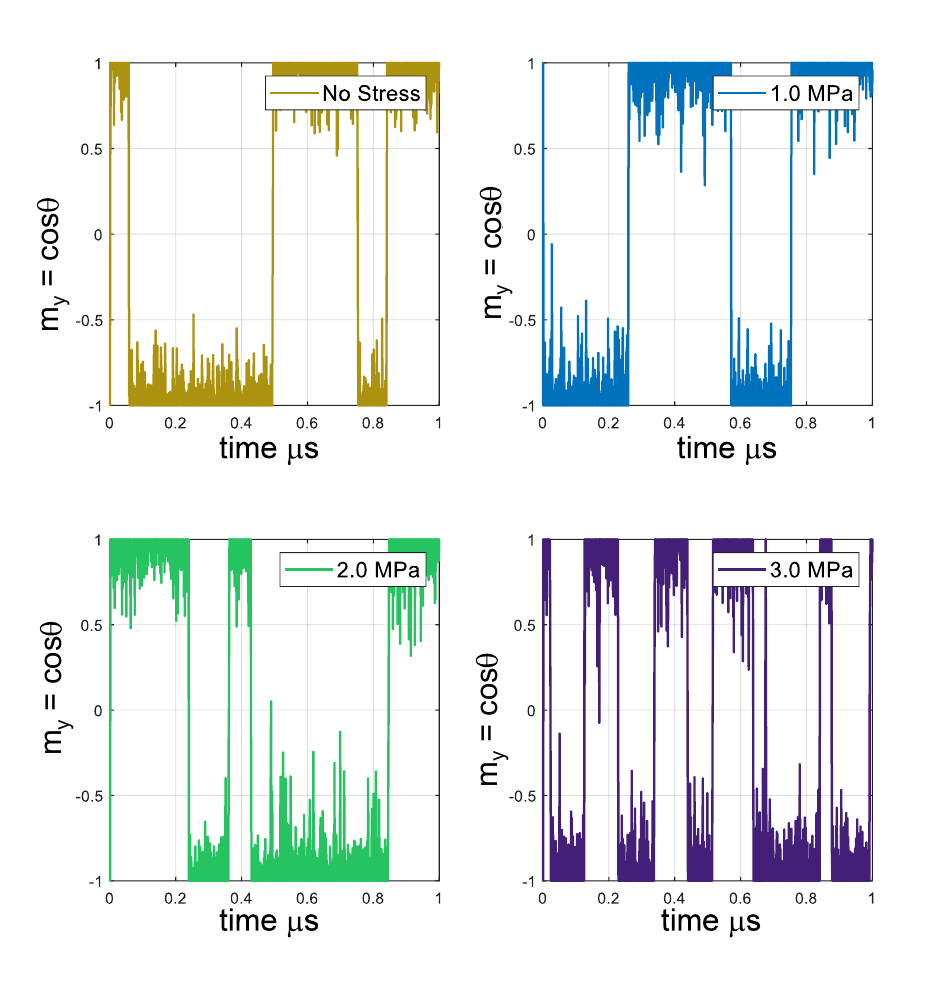}
    \includegraphics[width=0.46\textwidth]{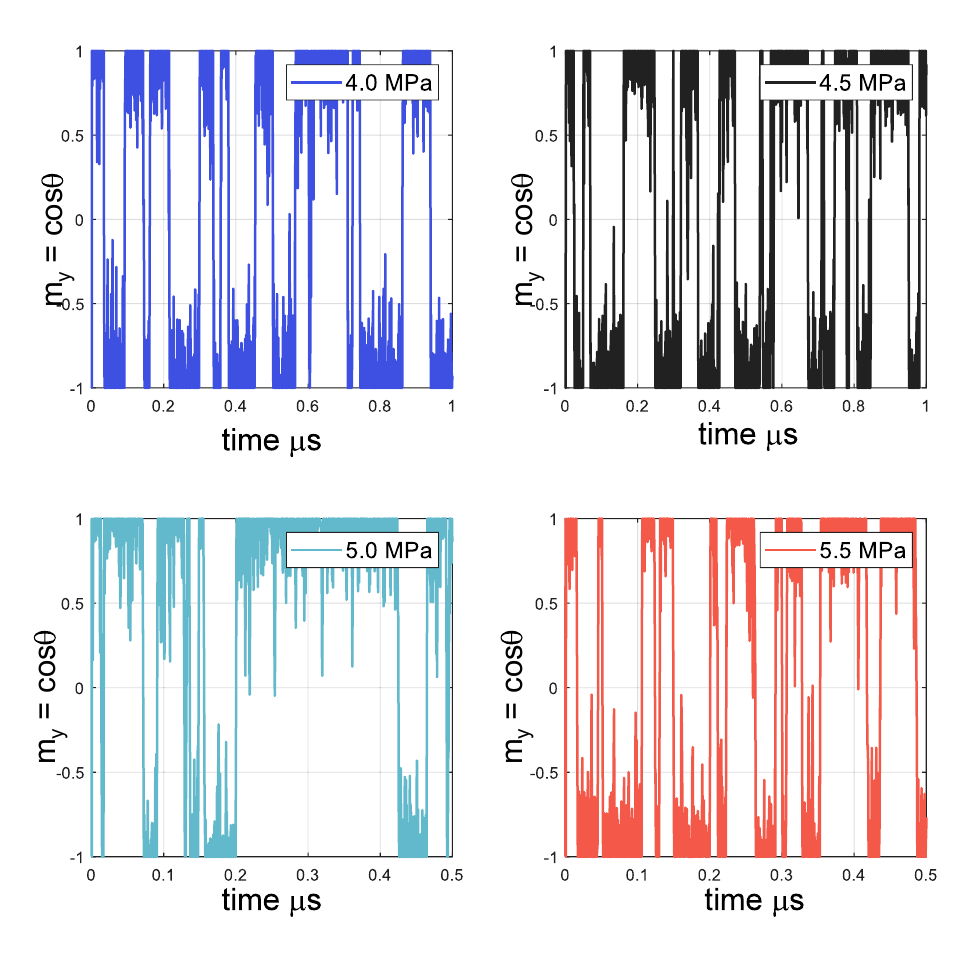}
    \centering
    \includegraphics[width=0.62\textwidth]{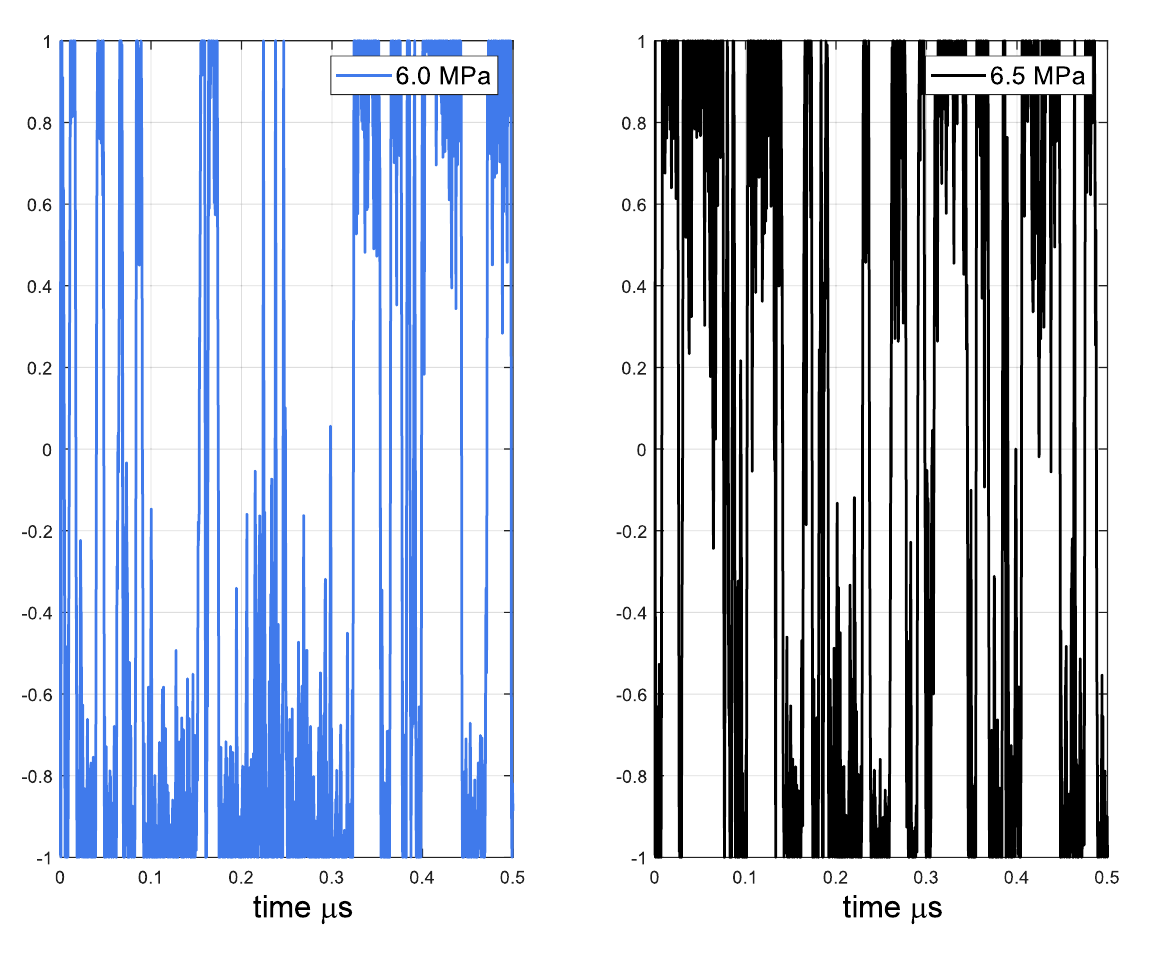}
    \caption{\small Temporal fluctuations in the magnetization component directed along the major axis of the nanomagnet $\left ( m_y \right )$ at different values of stress. Note that the behavior gradually transitions from BSN to ASN with increasing stress as the energy barrier within the nanomagnet is progressively depressed.}
    \label{fig:fluctuations}
\end{figure*}

Fig. \ref{fig:fluctuations} shows the time variations of the normalized magnetization component along the major axis of the nanomagnet, i.e. $m_y$ (which is also cos$\theta$) under different tensile stress. The magnetization is normalized to the saturation magnetization. Clearly under no stress, the behavior is that of a BSN where the magnetization fluctuates rail to rail and is mostly in the state +1 or -1, and not in any intermediate state. As we increase the stress (and depress the energy barrier), the behavior gradually transitions to that of an ASN wherein the magnetization visits all states between -1 and +1 with almost equal likelihood.

\subsection{Energy cost of reconfiguration}

We can make an order estimate of the energy cost of reconfiguration. This is the quantity $(1/2)CV_g^2$, where $C$ is the capacitance of the two gate pads in Fig. 3 and $V_g$ is the gate voltage needed to generate the required stress. We see from Fig. \ref{fig:fluctuations} that 6.5 MPa of stress is enough to reconfigure a BSN into an ASN. The gate voltage needed to generate a given stress $\sigma$ is given by $V_g \approx \sigma d /(Y d_{33})$, where $d_{33}$ is the diagonal element of the piezoelectric tensor and $d$ is the piezoelectric layer thickness. We will assume that the piezoelectric is PMN-PT whose reported $d_{33}$ value is 2500 pC/N \cite{shrout} and that $d$ = 300 nm. The Young's modulus of Co is 209 GPa. This will make the gate voltage needed to generate 6.5 MPa of stress = 3.6 mV. The capacitance $C$ of the two gate pads is $C$ = $2 \times \epsilon A/d$, where $\epsilon$ is the dielectric constant of PMN-PT = 4000$\times$8.854$\times$10$^{-12}$ F/m \cite{pramanik} and $A$ is the area of the gate pads = 100 nm $\times$ 100 nm. This makes $C$ = 2.4 fF. Hence the energy cost of reconfiguration $(1/2)CV_g^2$ is only $\sim$ 8.5$\times$10$^{-20}$ Joules, which is miniscule. The gate voltage of 3.6 mV is above the noise voltage at room temperature, which is $\sqrt{kT/C}$ = 1.3 mV ($kT$ = thermal energy). If more noise resilience is desired, one can increase the gate voltage beyond 3.6 mV to obtain the desired noise margin, while still dissipating negligible energy to reconfigure.

\section{Application Space for Dynamically Reconfigurable Stochastic Neurons}

Dynamic reconfigurability of the barrier height in a low barrier nanomagnet through precise voltage (strain) control  opens up some interesting possibilities in neuromorphic hardware fabrics. We list a few potential applications.

\subsection{Precision and Adaptive Annealing Control in Energy-based Computation}

One of the most important application of BSN is in solving binary optimization problems, specifically finding a bit string that minimizes a complex Boolean expression \cite{GloverQUBO}. Often such problems are in computationally intractable NP-Complete class and a Monte Carlo approach yields a computationally tractable probabilistic solution method \cite{Hochba}. Stochastic neurons can be used to build natural Monte Carlo hardware engines for solving such problems. The problem expression can be cast as an adjacency matrix of a network spanning over the bitstring basis set, and this adjacency matrix is then analogous to the Hamiltonian of an equivalent Ising network, where the bitstring represents the Ising spins \cite{BianDWAVE}. In a purely software implementation, an annealing process is simulated where the ``temperature'' of the system is slowly reduced and the system settles in its ground state which encodes the solution bitstring. 

In a hardware substrate, we directly build the Ising network and perform a physical annealing. Typically, such a hardware approach will use low barrier magnets (LBMs) and a current based scaling factor will control the ``inverse'' temperature \cite{CamsariPRX}. This current factor is global in nature and therefore scaling this current up cools the whole network \cite{Sutton}. In contrast, the approach presented here can seamlessly control the temperature via the control over the barrier height through a gate voltage rather than through current control. This is significantly easier to implement in a VLSI circuit.

Moreover, since this voltage control is directly over each individual neuron, completely arbitrary annealing schedules can be implemented with only polynomial increase in gate control circuit complexity and it brings {\it precision control} over the annealing approach, i.e. a screwdriver instead of a hammer.

\subsection{Control over Device-to-Device Variability and Data Retention Time}

\begin{figure}
    \centering
    \includegraphics[width=0.7\textwidth]{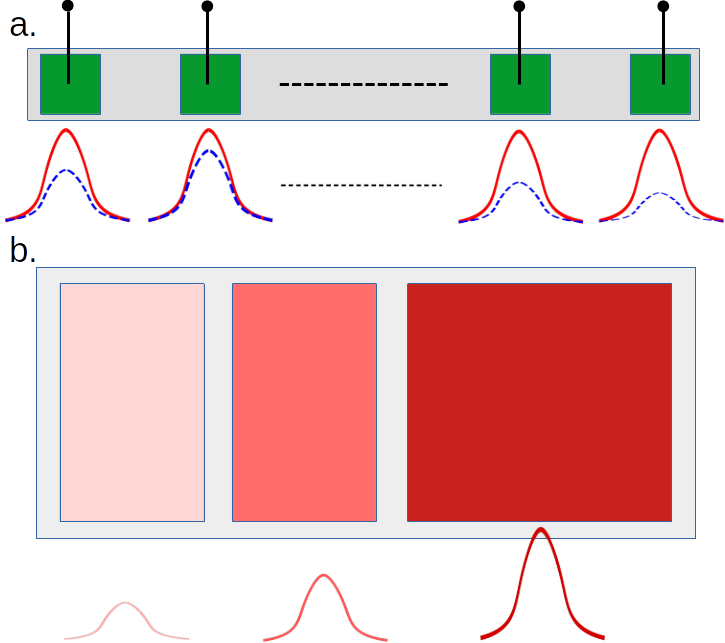}
    \caption{(a) Gate control over the barrier height (blue dashed line to red solid line) of each individual memory cell to equalize memory retention and hence reduce device-to-device variability in a memory array. (b) Gate control over sections of memory fabric to emulate memory hierarchy in terms of retention time-scales through barrier height modulation (color coded as light to dark red boxes and corresponding barrier heights), e.g. $\sim \mu s$, $\sim s$, $\sim$ years in a single integrated fabric}
    \label{fig:app2}
\end{figure}

The gate control of barrier height via strain can be also used to control variability in the barrier height in a large network. Deviation from the designed shape of a LBM due to lack of precision in the fabrication method can lead to significant variations in the natural barrier heights in LBM networks. It has been shown that in a network with large barrier height variability, significant error in computational results can be expected unless the compute process is exponentially long on the magnitude of the barrier height variability spread in the network, which in most practical applications is unacceptable from the perspectives of both energy cost and throughput \cite{Morshed}. Therefore, gate control of barrier height (fig. \ref{fig:app2} a) can enable both better energy efficiency and throughput.

On a different note, magnetic random access memory (MRAM) has long been touted as a solution to the memory hierarchy problem. However it's use has thus far been confined to replacement of solid-state non-volatile memory (NVM). The retention time in a magnetic memory cell scales as $exp(\Delta/kT)$, where $\Delta$ is the energy barrier height in the storage nanomagnet. The ability to dynamically control the barrier heights in certain sections of a large memory array may provide the ability to control memory retention times of sections of the memory fabric (fig. \ref{fig:app2} b). Since the vast majority of computing architecture and programming models are optimized to a memory hierarchy of speed/data retention rates, the ability to emulate the memory hierarchy in a dynamically reconfigurable fashion can open up intriguing possibilities of software defined hardware architectures, especially in the new age of heterogeneous integration and chiplet based compute fabric design.

\subsection{Control over Belief Uncertainty in Analog Stochastic Neurons}

\begin{figure}
    \centering
    \includegraphics[width=0.7\textwidth]{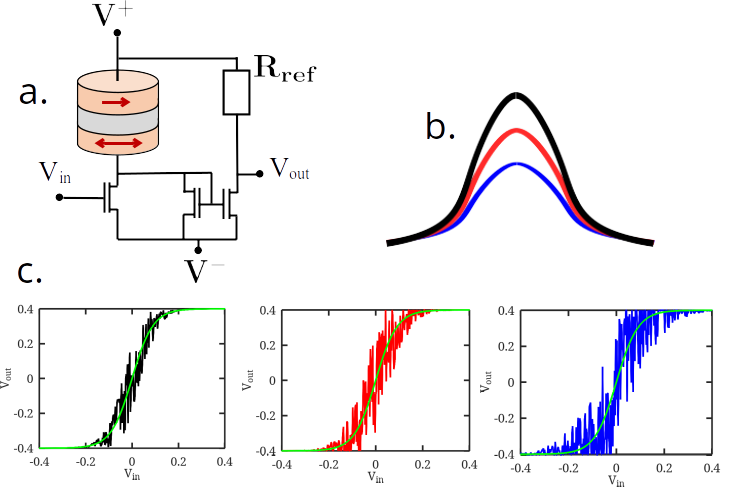}
    \caption{(a) Illustrative circuit design of an ASN cell with a stochastic MTJ (piezoelectric substrate omitted for clarity). (b) Gate control over the barrier height modulates the stochasticity. (c) Three example input-output characteristic curves of ASN transfer function (noisy signal is instantaneous output, green smooth signal is expected time averaged output) over three different barrier heights shows the degree of uncertainty on the belief of the ASN over its transfer function.}
    \label{fig:app3}
\end{figure}

In analog-stochastic-neurons the noise is banded around the sigmoidal transfer function and it's magnitude is maximal at the mid point of the transfer curve while minimal at the edges (fig. \ref{fig:app3} a) \cite{ASNMainPaper}, a feature which reflects the belief uncertainty of the neuron in its state over the transfer curve:

\begin{equation}
    V_{out}(t) = \tanh(\beta * V_{in}(t)) + \alpha(V_{in}(t))*V^0_{noise}(t)
\end{equation}

Where $\beta$, $\alpha$ are system transfer gain and noise profile function respectively, $*$ is the convolution operator, $V^0_{noise}$  is the normalized noise voltage (i.e. between $+V_{DD}/2$ and $-V_{DD}/2$), with $V_{DD}$ being the power supply voltage. Through detailed simulations we have observed that the empirical expression for $\alpha$ is given by the Gaussian profile:

\begin{equation}
    \alpha(V_{in}(t)) = \kappa\exp(-\frac{\nu V_{in}}{\sigma_{V_n}^2})
\end{equation}
where $\kappa$, $\nu$ are non-linear fitting functions dependent on the barrier height, whereas $\sigma_{V_n}$ is the standard deviation of the noise voltage profile, again a non-linear function of barrier height and the transfer gain of the neuron cell.

Voltage control over the noise through barrier height modulation (fig. \ref{fig:app3} b)  provides a clean way to control dynamically the magnitude of the belief and its spread (fig. \ref{fig:app3} c) . This can be used in applications such as reservoir computing in  continuous online-training mode where the network can reduce the noise during the training phase, while incorporating it during the inference stage \cite{MorshedToolsPaper}. This provides both robustness to the reservoir computing operation, i.e. the training can be made accurate, and the inference can be performed with significantly lower spread of the dynamic range of the observation model weight spread.

\section{Conclusion}

We showed that strain mediated control of the energy barrier height in a low barrier nanomagnet allows reconfiguring a BSN to an ASN and vice versa, thus allowing a multitude of tasks to be performed in the same substrate. We also discussed other applications of this modality.
These few applications are only a small subset of the potential functionalities that can be achieved in a gate controlled stochastic neuron compute fabric. The dynamic and precision control over the individual neurons through voltage control is well suited for conventional VLSI design methodologies and fabrication practices. Further exploration and development of this technology will provide a useful widget in the toolkit of the rapidly expanding discipline of hardware neuromorphics.

\pagebreak

\end{document}